\documentclass[12pt]{article}
\usepackage{latexsym,epsfig}
\usepackage{graphicx}
\usepackage{verbatim}

\def\ed{\end{document}}

\def\beq{\begin{eqnarray}}
\def\eq{\end{eqnarray}}
\def\beqn{\begin{eqnarray*}}
\def\eqn{\end{eqnarray*}}

\begin{document}

\title {Double seesaw mechanism  in a left-right symmetric model with TeV neutrinos}
\author{F.~M.~L.~de Almeida Jr., Y.~A.~Coutinho, 
J.~A.~Martins Sim\~oes, \\
A.~J.~Ramalho, L.~Ribeiro Pinto, S.~Wulck\\
Instituto de F\'isica \\
Universidade Federal do Rio de Janeiro\\
Rio de Janeiro, RJ, Brazil\\
M.~A.~B.~do Vale\\
Departamento de Ci\^encias Naturais\\
Universidade Federal de S\~ao Jo\~ao del Rei \\
S\~ao Jo\~ao del Rei, MG, Brazil }

\date{}

\maketitle

\begin{abstract}

A left-right symmetric model is discussed with new mirror fermions and a Higgs sector with two doublets and neutral scalar singlets. The seesaw mechanism is generalized, including not only neutrino masses but also charged  fermion masses. The spectrum of heavy neutrinos presents a second seesaw mass matrix and has neutrinos masses naturally in the TeV region. The model has very clear signatures for the new neutral vector gauge bosons. Two classes of models are discussed. New mirror neutrinos can be very light and a new $Z^{\prime}$ can be discriminated from other models by a very high invisible branching fraction. The other possibility is that mirror neutrinos can have masses naturally in the TeV region and can be produced through $Z^{\prime}$ decays into heavy neutrino pairs. Signatures and production processes for the model at the  LHC energy are also presented.

\end{abstract}



\section{Introduction}\
There is an increasing experimental evidence for neutrino oscillations coming from atmospheric, solar, reactor  and long-baseline accelerator neutrino experiments \cite{ALT, SCH, GON}. These results imply that neutrinos have small masses and large mixing parameters. Presumably these properties are the first manifestation of a new scale in nature. The flavour oscillations and the smallness of neutrino masses can be related to a simple property of neutrinos - they can be completely neutral and considered as Majorana particles. Majorana masses can arise from the seesaw mechanism  \cite{RAM} and the small masses can be related to the scale where lepton number is violated. This point opens important issues in particle physics, astrophysics and cosmology and can also be related to many extensions of the standard model. However, the canonical seesaw model is practically impossible to be experimentally verified. Very large masses and very small mixing parameters decouple heavy neutrinos from any feasible interactions, but models can be built to evade this situation. Based on Weinberg's argument \cite{WEI} that new physics must be contained in a dimension-5 operator, it is possible that neutrino masses can be related to other physical scales, below grand unification. If neutrino masses are not simply connected with the standard model Higgs scale, then we can expect that the charged fermion masses also have another origin.

A natural extension of the standard model is the left-right symmetric model \cite{JCP}, based on the group $SU(2)_L \otimes SU(2)_R \otimes U(1)_{(B-L)}$. The general idea is that parity is conserved at high energies and spontaneously broken in order to reproduce the standard model asymmetric interactions.  Many extensions of left-right symmetric models are possible. 

In this paper we will study a particular model \cite{NOS}: new mirror fermions \cite{RFO} are introduced  and related by a parity symmetry with the standard model fermions. A Higgs sector is considered with two doublets and neutral singlets. It was recently shown \cite{CHA, CHV} that we can built a scalar potential that gives a minimum consistent with low energy phenomenology. Singlet stable scalars are possible dark matter candidates \cite{DMO, KIM}. A $Z_2$ symmetry must be imposed to prevent decays. Singlet dark mass candidates with masses in the TeV region were recently investigated in ref. \cite{Ponton:2008zv}.
Our model will have new gauge interactions at a scale given by the breaking of $SU_{R}(2)$. We will explore the possibility that this scale is in the TeV region and will be accessible at the LHC. Parity will be broken at a much higher scale by the neutral singlet sector, as in the $D$-parity mechanism developed by Chang, Mohapatra and Parida  \cite{CMP}.
 
\section{The Model}\

Our Higgs sector has  two doublets $ \chi_{L} $ and $ \chi_{R} $  and  three singlets $S_{D}$, $S_{M_L}$  and $S_{M_R}$. Under  $ SU(2)_L \otimes SU(2)_R \otimes U(1)_{(B-L)}$ they transform as $  (1/2,\, 0, \, 1)$ , $(0, \, 1/2, \, 1) $ and $ (0, \, 0, \, 0)$ respectively.
Under $D$-parity they have the following transformations

\begin{equation}
\chi_{L} = \left(
\begin{array}{c}
\chi_{L}^{+} \\
\chi_{L}^{0} \\
\end{array}\right) \quad \buildrel {\rm D} \over \longleftrightarrow \quad
\chi_{R} = \left( 
\begin{array}{c}
\chi_{R}^{+} \\
\chi_{R}^{0} \\
\end{array}\right)\\
\end{equation}\

\begin{equation}
\chi_{L} \buildrel {\rm D} \over \longleftrightarrow \chi_{R},\quad
S_{M_L} \buildrel {\rm D} \over \longleftrightarrow - S_{M_R} \quad , \quad 
S_{D} \buildrel {\rm D} \over \longleftrightarrow S_{D} \nonumber
\end{equation}\

After spontaneous symmetry breaking the scalar fields have vacuum expectation parameters accordingly to

\begin{eqnarray}
\langle \chi_{L} \rangle = \left(
\begin{array}{c}
0 \\
v_{L} \\
\end{array}\right), \quad
\langle \chi_{R} \rangle = \left( 
\begin{array}{c}
0 \\
v_{R} \\
\end{array}\right), \quad
\langle S_{D} \rangle = s_{D} \quad ,  \quad
\langle S_{M_{L,R}} \rangle = s_{M_{L,R}}.
\end{eqnarray}\

 The two charged vector bosons will have masses proportional to $v_L$ and $v_R$ and will not mix at tree level. This is a consequence of our choice for the Higgs sector. As the scalar singlets can be mixed, this will reintroduce mixing through loop corrections that must be small if the new scalar singlets have large masses. We must identify $ v_L = v_{Fermi}$ and from the absence of right-handed currents, $ v_R > v_L$. The neutral vector gauge bosons will mix in a simple way and the present bound on $w = v_{L}/v_{R}$ implies \cite{NOS} $v_{R} > 30 \, v_{L}$. The massive neutral gauge bosons will have masses (see ref. [7] for details)

\begin{eqnarray}\label{autovaloresR}
M^2_Z &=& \frac{1}{4} \frac{v^2_L \,g^2_L}{\cos^2\theta_W} \left( 1 - \omega^2 \sin^4 \beta \right) \nonumber \\
M^2_{Z^{\prime}} &=& \frac{1}{4} v^2_R \, g^2_L \, \tan^2 \theta_W \tan^2 \beta \left( 1 + \frac{\omega^2 \sin^2 2\beta}{4 \sin^2\theta_W} \right).
\end{eqnarray}\

The mixing angles are given by

\begin{eqnarray}\label{g}
\sin^2\theta_W &=& \frac{g^2_R \, g^2}{g_L^2 \, g_R^2 + g_L^2 \, g^2 + g_R^2 \, g^2},  \nonumber \\
\sin^2\beta & = & \frac{g^2}{g_R^2 + g^2}. \quad  \nonumber \\
\end{eqnarray}\

The fermion content of the model is given by 

\begin{equation}
l_{e \,L}=\left(
\begin{array}{c}
\nu_{e} \\
e \\
\end{array}\right)_L,\
\nu_{e \,R},\
e_R \quad ; \quad
L_{e \, R}=\left(
\begin{array}{c}
N_{e} \\
E \\
\end{array}\right)_R,\
N_{e \,L},\
E_L
\end{equation}\,

and similar relations for the other fermionic families. As we are including new chiral fermions for each lepton and quark family, the anomaly cancelation has the same pattern as in the standard model. Under $D$-parity they transform as:
\begin{eqnarray*}
l_{e \,L} \quad \buildrel {\rm D} \over \longleftrightarrow \quad L_{e \, R}, \qquad \nu_{e \,R} \quad \buildrel {\rm D} \over \longleftrightarrow \quad N_{e \, L},  \qquad e_R \quad \buildrel {\rm D} \over \longleftrightarrow \quad E_{L}.
\end{eqnarray*}

Fermion masses are generated from the following $ SU(2)_L \otimes SU(2)_R \otimes U(1)_{(B-L)} \otimes D $   invariant Lagrangian:

\begin{eqnarray}
\cal{L} &=& \lambda \left[\overline{l_{e \,L}} \,  \chi_L \, e_R + \overline{L_{e \,R}} \, \chi_R  \, E_L +\overline{l_{e \,L}}\, \tilde{\chi_L} \, \nu_{e \,R} + \overline{L_{e \,R}} \,  \tilde{\chi_R} \,  N_{e \,L} \right]  \nonumber \\
& + & \lambda^{\prime} \left[\overline{l_{e \,L}} \, \tilde{\chi_L} \, N^c_{e \,L} + \overline{L_{e \,R}} \, \tilde{\chi_R} \, \nu^c_{e \,R} \right] 
 + \lambda^{\prime \prime} \left[S_{M_R} \, \overline{\nu^c_{e \,R}} \, \nu_{e \,R} - S_{M_L} \, \overline{N^c_{e \,L}} \, N_{e \,L}\right] \nonumber\\ 
& + & g^{\prime} \, S_D  \, \overline{\nu_{e \,R}} \, N_{e \,L} + g^{\prime \prime} \, S_D \overline{e_R} \, E_L. 
\end{eqnarray}\

The charged fermion spectrum is the same as defined in ref. 7 . But as the  singlet scalar sector is not the same, the neutrino spectrum will be quite different. The charged fermion mass spectrum can be easily obtained by including the three families with new Higgs couplings given by $\lambda_i$, $g_i$  where  $i= e, \, \mu, \, \tau$. These new parameters will give different masses for each family. The charged light and heavy fermion masses come from a seesaw mechanism \cite{NOS} and their values are 

\begin{equation}
m_{i}=\frac{\lambda_{i}^2}{g^{\prime \prime}_{i}}  \frac{v_{L} \, v_{R}}{  s_{D}} \qquad {\textnormal and}  \qquad M_{i}=g^{\prime \prime}_{i} \,  s_{D},
\end{equation}

 We see that the charged mass spectrum is not simply given by the Fermi scale $v_L$ as in the standard model but also involves the new $v_R$ and $s_D$ scales. Let us consider the electron mass. If we fix the $v_R $ scale at the LHC energies   $v_R \simeq 10^{4} ~$ GeV  and for the  choice $\displaystyle{\frac{\lambda_{1}^2}{g^{\prime \prime}_{1}}\simeq 1}$ , we have for the  $s_D $ scale the value

\begin{equation}
 s_{D} \simeq 10^{ 10} \,\,\, {\textnormal GeV}.
\end{equation}\

A first important consequence  of the model is the fact that the electron mass (and the other charged fermion masses) also comes from the seesaw mechanism, and that the scalar singlet scale is of the order of the Peccei-Quinn scale to solve the strong CP problem \cite{CAR}. Our model has no strong CP violation before symmetry breaking.  After symmetry breaking it is possible to estimate the strong CP violation, but we must then introduce new parameters in the model. The other charged fermion masses can be generated in a similar way.

The interactions in Lagrangian (7) are radiatively stable, but we can not prevent the  scalar masses from acquiring large values from higher order corrections. The reason for this is very simple: we are "doubling" the standard model in a new right-handed sector. So the problem of hierarchy will appear in the same way as it does in the standard model. The solution for this problem must come from new symmetries as is the case for supersymmetry or from new conditions (fine tuning) among the couplings in the new interactions. 

A related question is that mixing could be induced by loop contributions and there are strong bounds on the mixing parameters\cite{Langacker:1989xa}. The question of $W_R /W_L $ mixing is treated in ref. \cite{BRA} including one-loop effects of new right-handed charged currents for the known quarks and leptons.  Our model has a left-right symmetry with the inclusion of new mirror fermions. The quark and lepton usual right handed singlets remains as singlets in our model and do not have new right-handed charged current couplings. So, the one-loop contribution to $W_R /W_L $ mixing shown in Figure 1 of ref. \cite{BRA}  does not exist in our model. Another possible source of mixing discussed in the previous reference, comes from the Higgs bi-doublets fields which belong to both left and right sectors. But again, our model has no such bi-doublet fields. A third, indirect source of $W_R /W_L $ mixing could came from fermion mixing. Here again, our choice for the Higgs sector  allows important mixing only for the neutrino sector, not for charged fermions. This will not contribute to $W_R /W_L $ mixing. So, in our model there are no one-loop contributions to $W_R /W_L $ mixing. This result agrees with the more general treatment of the Higgs sector done in ref. \cite{MAS}. It is also well known that one-loop fermionic contribution to $\gamma - Z $ mixing must be finite. The usual standard model fermions satisfy this condition. As  mirror fermions have opposite helicities relative to their standard model partners, they can be treated in the same way. Here again there will be no infinities due to one-loop corrections. This completes the proof that there will be no one-loop infinities to gauge boson mixing in the our model.

 In order to fully demonstrate that there are no new large terms in the model, one should also demonstrate that there is no hierarquy problem. This can only be achieved if new properties are imposed, not only in our model, but possibly also in all left-right models.
We point out that our model is  based on a renormalizable, anomaly-free approach,
with no tree level ambiguities and no one-loop high corrections. We will not develop further on these points but we stress that in ref. [9,10] it was shown that the scalar potential has a minimum and that we can have naturally the condition $ v_R > v_L $.  

\section{Neutral lepton masses}\
\par
For the neutral lepton masses we can rewrite the neutral sector in terms of the Majorana fields:

\begin{eqnarray}
\chi_{\nu} & = & \nu_L + \nu_L^c  \nonumber \\
w_N & = & N_R + N_R^c \nonumber \\
\chi_N & = & N_L + N_L^c \nonumber \\
w_{\nu} & = & \nu_R + \nu_R^c.
\end{eqnarray}\

 In matrix form, the neutral lepton Lagrangian is given by

\begin{eqnarray}
&& \mathcal{L}^{NC}_L = \overline{\xi} \, M_n \, \xi = \nonumber \\
 && \left(\overline{\chi}_{\nu} \,\, \overline{w}_N \,\, \overline{\chi}_N \,\,
\overline{w}_{\nu} \right)
\left(\begin{array}{cccc}
0 & 0 & \lambda^{\prime} v_L &  \lambda \displaystyle{\frac{v_L}{2}}\\
0 & 0 & \lambda \displaystyle{\frac{v_R}{2}} & \lambda^{\prime} v_R \\
 \lambda^{\prime}v_L & \lambda \displaystyle{\frac{v_R}{2}} & -  \lambda^{\prime \prime} s_{M_L} &  \displaystyle{ g^{\prime} \frac{s_D}{2}} \\
 \lambda \displaystyle{\frac{v_L}{2}} & \lambda^{\prime} v_R &  g^{\prime} \displaystyle{\frac{s_D}{2}} &   \lambda^{\prime \prime} s_{M_R} 
\end{array} \right)
\left(\begin{array}{c}
\chi_{\nu}\\
w_N\\
\chi_N\\
w_{\nu}
\end{array} \right)
\end{eqnarray}\

Defining the block matrices,

\begin{eqnarray} 
M_{LR}=\left(\begin{array}{cc}
  \lambda^{\prime} v_L &  \lambda  \displaystyle{\frac{v_L}{2}}\\
 \lambda \displaystyle{\frac{ v_R}{2}} &  \lambda^{\prime}  v_R \\
\end{array} \right)  \quad
M_S=\left(\begin{array}{cc}
-   \lambda^{\prime \prime} s_{M_L} &  g^{\prime} \displaystyle{\frac{s_D}{2}}\\
 g^{\prime}\displaystyle{\frac{ s_D}{2}} &   \lambda^{\prime \prime}  s_{M_R} \\
\end{array} \right)
\end{eqnarray} \

then the block diagonalization procedure implies

\begin{equation}
M^{(light)}\simeq -M_{LR}^t \, M_S^{-1} \, M_{LR}.
\end{equation}\

For the heavy mass matrix we have

\begin{eqnarray} \label{massapesada}
M^{(heavy)}_{n}=\left(\begin{array}{cc}
- \lambda^{\prime \prime} s_{M_L} &  g^{\prime} \displaystyle{\frac{s_D}{2}}\\
 g^{\prime} \displaystyle{\frac{s_D}{2}} & \lambda^{\prime \prime}  s_{M_R} \\
\end{array} \right).
\end{eqnarray} \

 In order to proceed with the generalization to three families, we must consider the differences between the charged and neutral sectors. Mixing between families in the charged sector is phenomenologically disfavored. But as we have neutrino singlets given by $\nu_R $ and $N_L$ that are completely neutral, they can mix families in different ways. So we  have to choose suitable candidates for the textures in the coupling matrices. One simple choice is to consider that all neutrino fields are physically equivalent and to take all their Yukawa couplings as equal. In this case we will have only one  different coupling per generation for each term in the preceeding matrices. The exact neutrino mass and mixing angles will depend on the texture hypothesis, but the important point in our model is that the block structure of the neutrino mass spectrum will not depend on any texture hypothesis. Thus the size of neutrino masses can be estimated.

The order of magnitude of the "light" neutrino mass spectrum is then 

\begin{eqnarray}
m_{\nu_1}= \frac{\lambda}{4}\frac{v_L^2}{s_{M_R}}, \qquad m_{\nu_2}=\frac{\lambda}{4}\frac{v_L^2}{s_{M_R}},\qquad m_{\nu_3}\simeq \frac{33 \lambda}{4} \frac{v_L^2}{s_{M_R}}  \nonumber \\
m_{N_1}= \frac{\lambda}{4}\frac{v_R^2}{s_{M_R}},\qquad m_{N_2}=\frac{\lambda}{4}\frac{v_R^2}{s_{M_R}},\qquad m_{N_3}\simeq \frac{37 \lambda}{4}\frac{v_R^2}{s_{M_R}}.
\end{eqnarray}\
 The first three masses $m_{\nu_i}$ can be identified with the standard neutrinos masses and the other three $m_{N_i}$ will be discussed in the next section.

For the Higgs doublets we have $ v_R >> v_L $. It is also reasonable to take $ s_{M_R} >> s_{M_L} $ and the heavy mass matrix becomes
\begin{eqnarray}
M^{(heavy)}_{n}=\left(\begin{array}{cc}
0 &  g^{\prime} \displaystyle{\frac{s_D}{2}}\\
 g^{\prime} \displaystyle{\frac{s_D}{2}} & \lambda^{\prime \prime}  s_{M_R} \\
\end{array} \right).
\end{eqnarray} \

If $ s_{M_R} >> s_D $ we have a repetition of the seesaw mechanism in the heavy mass matrix. We call this property the "double seesaw mechanism"
\cite{DSS}.

The mass eigenvalues are

\begin{eqnarray}
M_1 = \lambda^{\prime \prime} s_{M_R}  \nonumber \\
M_2= \frac {g^{\prime^2}}{\lambda^{\prime \prime}} \frac{s_D^2}{s_{M_R}}.
\end{eqnarray}\
If we take $ s_{M_R}\simeq 10^{16}$ GeV, $s_D \simeq 10^{10} $ GeV and $\displaystyle{\frac {g^{\prime^2}}{\lambda^{\prime \prime}} \simeq 0.1}$, then we have a heavy neutrino with a mass $ M_2 \simeq 1 $ TeV.

\section{The New Neutral Currents}\
In order to identify the neutrino spectrum with the interacting states we must first discuss the neutral current content of the model.
 After rotation of the neutral vector gauge bosons \cite{NOS}, the neutral current interactions are given by

\begin{eqnarray}
\mathcal{L^{(\nu ,N)}}&=&-J^{(\nu ,N)}_{\mu} \, Z^{\mu}-J^{\prime (\nu ,N)}_{\mu}Z^{\prime \mu}= \nonumber \\
& = &-\frac{g_L}{2 \cos\theta_W}\left[(1-w^2 \sin^4 \beta )\overline{\nu_L}\gamma^{\mu}\nu_L- w^2 \sin^2\beta \overline{N_R}\gamma^{\mu} \, N_R \right]Z_{\mu}  \nonumber \\
& - & \frac{1}{2}g_L \tan\theta_W \tan\beta \left[\left(1 + w^2 \frac{\sin^2\beta \cos^2\beta}{\sin^2 \theta_W}\right) \overline{\nu_L}\gamma^{\mu} \nu_L \right. \nonumber \\
 & + & \left. \frac{1}{\sin^2 \beta} \overline{N_R}\gamma^{\mu}N_R \right]Z^{\prime}_{\mu}.
\end{eqnarray}\

In the Majorana basis

\begin{eqnarray}
\mathcal{L^{(\nu ,N)}}&=&-\frac{g_L}{2 \cos\theta_W} \left[(1-w^2 \sin^4 \beta ) \overline{\chi_{\nu }}\gamma_{\mu} \frac{(1 - \gamma_5)}{2} \chi_{\nu} \right. + {}  \nonumber \\
& - & \left. w^2 \sin^2\beta \sum_{i}^n \overline{w_{N}}\gamma_{\mu}\frac{(1+\gamma_5)}{2}w_{N} \right] Z^{\mu} {} \nonumber \\
& - & \frac{1}{2} g_L \tan\theta_W \tan\beta \left[(1 + w^2 \frac{\sin^2\beta \cos^2\beta}{\sin^2 \theta_W})\overline{\chi_{\nu}}\gamma_{\mu}\frac{(1- \gamma_5)}{2}\chi_{\nu } \right. + {} \nonumber \\
& + & \left. \frac{1}{\sin^{2}\beta} \overline{w_{N}}\gamma_{\mu}\frac{(1+\gamma_5)}{2}w_{N} \right] Z^{\prime \mu}.
\end{eqnarray}
\par

From these neutrino-neutral gauge boson interactions, and from the neutrino mass spectrum developed in the previous sections we can now proceed to identify the neutrinos. The generalization to three families is straightforward. From the $J^{(\nu ,N)}_{\mu}Z^{\mu}$ 
term we must identify the $ \nu_{L_i} $ (or the $ \chi_{\nu_i} $) fields as the three standard model neutrinos. They have masses of the order of $v_L^2 /s_{M_R}$ as in the usual seesaw model.
The new $ N_{R_i}$ ( or the $w_{N_i}$) are also coupled to the standard model $Z$ gauge boson, but this coupling is strongly suppressed by a factor $ w^2  < 10^{-3}$.  The light-to-heavy neutrino mixing is also very  small and can be neglected. In the light matrix we can reproduce the presently known mixing parameters \cite{NOS}. The heavy neutrino mixing will involve the new heavy Majorana masses and phases. This can lead to different phenomenological consequences. In this paper we will consider only two limiting cases: very light and heavy $N_{R_i}$ states.

\vfill\eject
{\bf Model A}  

The three $ N_{R_i}$ states are light, with masses of the order of $v_R^2 /s_{M_R}$. In this case the heavy Majorana states have very high masses and will be out of reach for any feasible experimental detection. But the model has a very distinctive signature \cite{ERL}, the new heavy gauge boson $ Z' $ is coupled with the standard neutrinos and also with the new light states. As these couplings are given by a $1/ \sin^2{\beta}$ term , the new gauge boson will have a high invisible branching ratio that  will be discussed  in the next section.

The constraints on neutrino masses come from cosmological considerations related to typical bounds on the universe mass density and its lifetime.
For neutrinos below $\simeq 1$ MeV the limit on  masses for Majorana type neutrinos is \cite{TUR}:

\begin{equation}
\sum_{\nu} m_{\nu}\leq 100 \, \Omega_{\nu}  {\textnormal h} ^2 \,\,\, {\textnormal eV} \simeq 1 \,\, {\textnormal eV}
\end{equation}

where $\Omega_{\nu}$ is the neutrino contribution to the cosmological density parameter, $\Omega$ and the factor $ {\textnormal h} ^2$ measures the uncertainty in the determination of the present value Hubble parameter, and the factor $\Omega \,  {\textnormal h} ^2$ is known to be smaller than 1.

From eq.(15), the sum of neutrino masses must satisfy 
\begin{equation}
\sum_i^6 m_i \leq 10 \lambda \frac{v_R^2}{s_M},
\end{equation}
so that the cosmological criterium  is verified if
\begin{eqnarray}
\lambda \frac{v_R^2}{s_M}\leq 10 \, \Omega_{\nu} \,\, {\textnormal h} ^2eV.
\end{eqnarray}

 This is a very interesting constraint since it can be used as an upper bound on $v_R$. If  the breaking scale $s_M$ is to be  $s_M \simeq 10^{16}$, $ \lambda \simeq 0.1$, then we have   $v_R < 10$ TeV. These new neutrinos interactions with the standard model $Z$ gauge boson are suppressed by a factor $w^2 \sin^2\beta $ in equations (18, 19) and could decay as $ N \rightarrow \nu \,+  \gamma$.

\vskip 1cm
{\bf Model B} 
 
The $ N_{R_i}$ states are heavy. As we have shown, the doubled seesaw mechanism can give masses for these neutrinos naturally at the TeV scale and this can be experimentally accessible at the LHC energies \cite{ALM}. For heavy neutrinos with masses up to $100-200$ GeV, the main  production mechanism is through $ W \rightarrow N + \ell $ with $ \ell = e, \mu, \tau$. However this mechanism implies a restrictive bound on the detection of heavy neutrino masses at LHC energies. The possibility of detecting heavy Majorana neutrinos with masses in the TeV region has been proposed as a test for many models \cite{ROD, NEW}. In our case, the production mechanism will be through $ Z^{\prime} \rightarrow N + N $ and heavy neutrino masses could be as high as $ M_N \simeq M_{Z^{\prime}}/2 $. Single heavy $N$ production through $Z^{\prime} \rightarrow N + \nu $ is strongly suppressed. We have three heavy Majorana neutrinos in the TeV region. In this paper we will consider only the lightest of these states.
Leptogenesis could be achieved by the decay of this lightest Majorana neutrino before the spontaneous breaking of the scalars fields. The coupling of this Majorana neutrino with the electron and the W gauge boson is bounded from neutrinoless double beta decay and the current limit is 
\cite{AAL}

\begin{equation}
\sin^2\theta_{eNW}  < 5 \times 10^{-8} \times M_N  {\textnormal \,\, (GeV)}
\end{equation}

\section{Signatures at the LHC}\

The new forthcoming data from the CERN Large Hadron Collider (LHC) will allow testing new physics at the TeV scale. In the left-right symmetric model the lightest particles are the new neutrino states and the most clear signal will be given by the new neutral gauge boson $Z^{\prime}$. A new neutral gauge boson is predicted by many extended models and one of the main goals of the LHC will be to test these possibilities \cite{ERL}. One of the cleanest signal for a new $Z^{\prime}$ is the process $p + p \longrightarrow Z^{\prime} + X \rightarrow  l^{+}+ l^{-}+ X$ with $l = e, \mu$. In Figure 1 we show the total cross section and number of events for an integrated annual luminosity of $ 100$ fb$^{-1}$  and a center of mass energy of $14$ TeV. We can estimate an upper bound on $M_{Z'}$ around $4.5$ TeV. We applied the following cuts on the final fermions: $\vert\eta\vert \leq 2,5$ since the LHC detectors have better tracking resolution in this $\eta$ range, 
$ M_{Z'} - 5 \Gamma_{Z^{\prime}} < M_{l^- l^+} < M_{Z^{\prime}} + 5 \Gamma_{Z^{\prime}}$ to avoid the standard model background and an energy cut in the final state charged leptons $ E_l > 5$ GeV. The model B total cross section for this channel is not very sensitive for the heavy Majorana mass $M_N$, hence we have drawn a single curve for this model with $M_N = 450$ GeV.

\begin{figure} 
\begin{center}
\includegraphics[width=.6\textwidth]{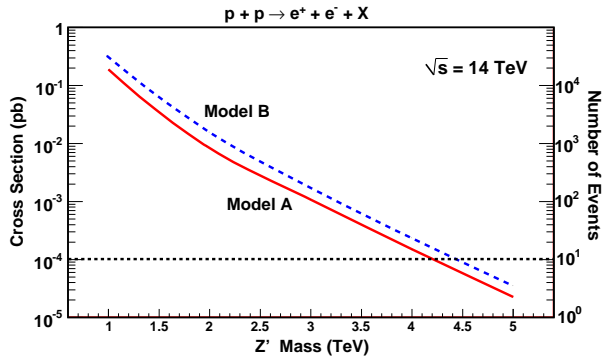}
\caption{Total cross section and number of events versus $M_{Z^{\prime}}$ in the process $p + p  \rightarrow  l^{+}+ l^{-}+ X$.}
\end{center} 
\end{figure}

In Figure 2 we show the peak difference between the two models and the enhancement in the total cross section due to the $Z^{\prime}$ production. The peak for model A is higher than the peak for model B.  The total width $\Gamma_{Z^{\prime}}$ is different for both models, as calculated in Tables 1 and 2.

\begin{figure} 
\begin{center}

\includegraphics[width=.6\textwidth]{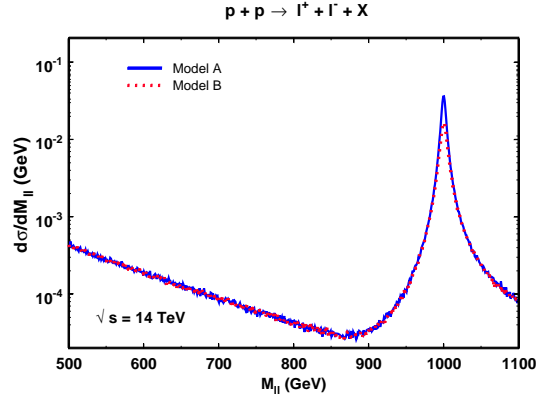}
\caption{Invariant mass distribution for the process $p + p \longrightarrow Z^{\prime} \rightarrow  l^{+} + l^{-} + X$ considering $M_{Z^{\prime}}= 1 $ TeV in models A and B.}
\end{center} 
\end{figure}

One of the simplest extensions of the standard model is the inclusion of singlet right-handed neutrino states (type-I seesaw models). In the mirror model proposed in this paper, the number of neutrino states is twice the number of neutrinos of the usual type-I seesaw model. If the new neutrino states are light (compared with the $Z^{\prime}$ mass) we will have a unique and very clear signature (model A): a high invisible branching ratio for the new possible $Z^{\prime}$. This is shown in Table 1, where we have shown the separated contribution  of the standard light neutrinos (which are also coupled to the Z) and the new light neutrinos $N_i$. They will all contribute to a very large decay $ Z^{\prime} \rightarrow {\textnormal invisible}$, of more than $ 43 \%$. In this case, the remaining Majorana states will be extremely heavy and out of reach even for the LHC. The new light neutrino states could decay radiatively into the standard neutrinos and this decay will involve new mixing parameters.

\begin{table}[ht]\label{Dachshund}
\begin{footnotesize}
\begin{center}
\begin{tabular}{||c|c|c||}
     \hline
\hline
&    &          \\ 
&  $M_{Z^{\prime}}= 1$ TeV &  $M_{Z^{\prime}}= 2$ TeV   \\
&    &          \\
&  $\Gamma_{Z^{\prime}}= 11.7$ GeV &  $\Gamma_{Z^{\prime}}= 23.4$ GeV   \\
\hline
&    &      \\
$Z^{\prime} \rightarrow \Sigma_i \, \bar \nu_i \nu_i$ & $4.5\%$ &
$4.5\%$  \\
&      &   \\ \hline
\hline
&     &     \\
$Z^{\prime} \rightarrow \Sigma_i \, \bar l_i l_i$ & $23.1\%$
& $23.1\%$  \\
&    &    \\ \hline
\hline
&     &        \\
$Z^{\prime} \rightarrow \Sigma_i N_i N_i $ &  
$39\%$  &  $38\%$  \\
&      &      \\ \hline
\hline
&       &   \\
$Z^{\prime} \rightarrow \Sigma_i \,\bar q_i q_i$  &   $33.7\%$  &
$33.8\%$  \\
&     &     \\  \hline
\hline
&     &        \\
$Z^{\prime} \rightarrow \Sigma_i \nu_i N_i $ &  
$0\%$  &  $0\%$  \\
&      &      \\ \hline
\hline

\end{tabular}
\end{center}
\end{footnotesize}
\caption{The $Z^{\prime}$ branching-ratios and widths in model A.}
\end{table}

We now turn our attention to model B. As we have shown in the previous sections, the double seesaw mechanism proposed here gives naturally  Majorana neutrino masses in the TeV region. We have three Majorana neutrinos in this region and the new ${Z^\prime}$ can decay in these states up to the kinematical limit $M_{N_i} = M_{Z^\prime}/2$. In model B we have considered only one of these states, the lightest one, with equal couplings to all charged leptons and W. The other states could be easily included. The new ${Z^\prime}$ branching fractions are shown in Table 2.

\begin{table}[ht]\label{Chihuahua}
\begin{footnotesize}
\begin{center}
\begin{tabular}{||c|c|c||}
     \hline
\hline
&    &          \\ 
&  $M_{Z^{\prime}}= 1$ TeV &  $M_{Z^{\prime}}= 2$ TeV   \\
&    &          \\
&  $\Gamma_{Z^{\prime}}= 7.4$ GeV &  $\Gamma_{Z^{\prime}}= 15.4$ GeV   \\
\hline
&    &      \\
$Z^{\prime} \rightarrow\Sigma_i \, \bar \nu_i \nu_i$ & $7.2\%$ &
$6.9\%$  \\
&      &   \\ \hline
\hline
&     &     \\
$Z^{\prime} \rightarrow \Sigma_i \,\bar l_i l_i$ & $36\%$
& $36\%$  \\
&    &    \\ \hline
\hline
&     &        \\
$Z^{\prime} \rightarrow N N $ &  
$6.1\%$  &  $6.4\%$  \\
&      &      \\ \hline
\hline
&     &        \\
$Z^{\prime} \rightarrow \Sigma_i \, \bar \nu_i N (N \nu_i) $ &  
$1.4 \times 10^{-3}\%$  &  $1.4 \times 10^{-3}\%$  \\
&      &      \\ \hline
\hline
&       &   \\
$Z \rightarrow \Sigma_i \,\bar q_i q_i$  &   $50.7\%$  &
$50.7\%$  \\
&     &     \\  \hline
\hline

\end{tabular}
\end{center}
\end{footnotesize}
\caption{The $Z^{\prime}$ branching-ratios and widths in model B.}
\end{table}

The most important signature for heavy Majorana neutrinos is the lepton number violating decay $ N \rightarrow l^{\pm} + W^{\mp} $ where $ l = e,\, \mu $. For heavy Majorana neutrinos with masses above $200$ GeV,  the heavy neutrino production and decay will be $p + p \longrightarrow Z^{\prime} \rightarrow N + N + X \rightarrow  l_i +l_j + W + W + X$. The final state pair of leptons $ l_i + l_j $ has lepton number violation through equal charges and/or electron and muon combinations.

\eject
\begin{figure} 
\begin{center}
\includegraphics[width=.6\textwidth]{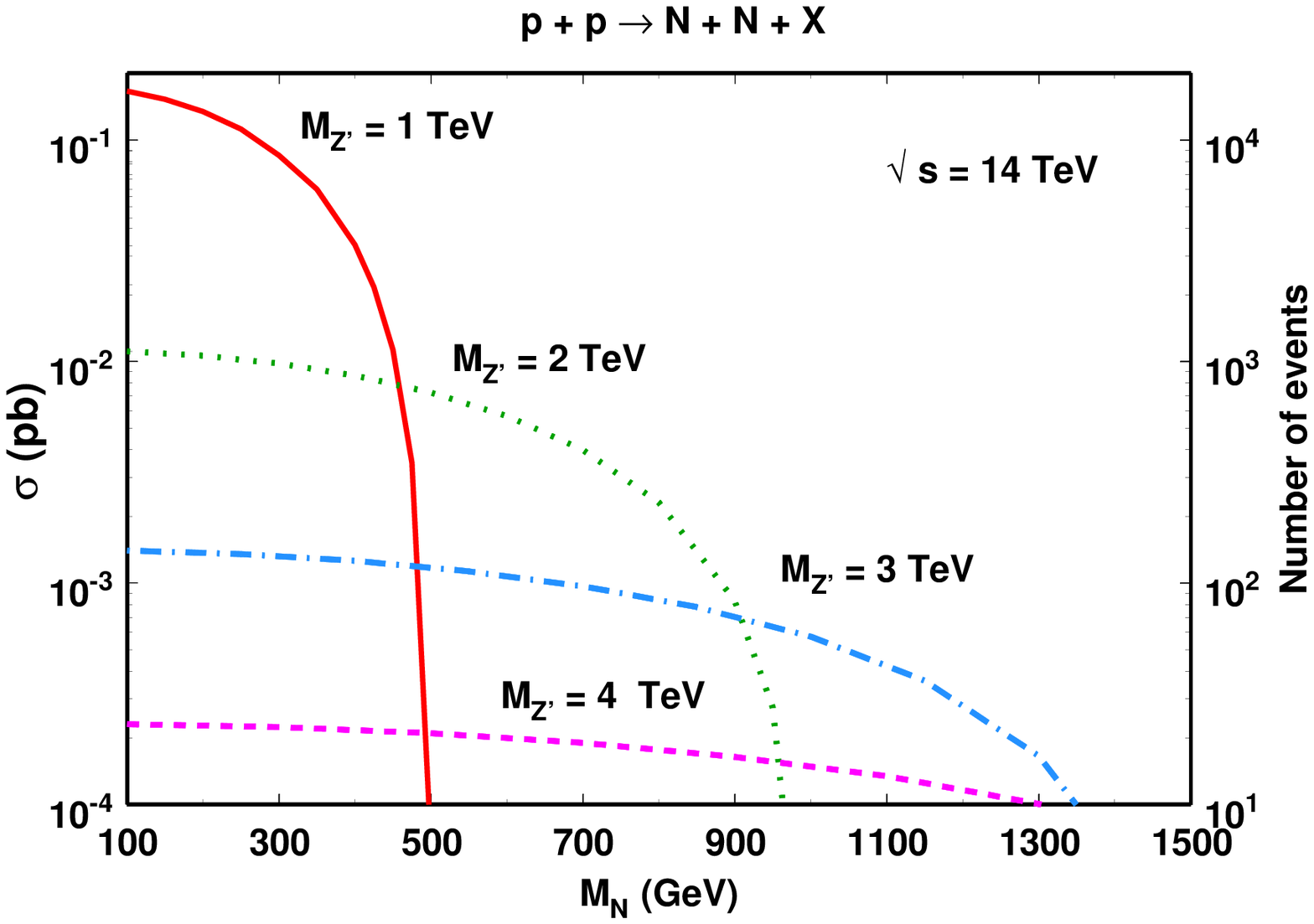}
\caption{Total cross section for the process $p + p \longrightarrow Z^{\prime} \rightarrow  N + N + X$ considering several values for $M_{Z^{\prime}} $ in model B.}
\end{center} 
\end{figure}

\section{Conclusion}
 In the present work we propose a model for the parity spontaneous breaking. New mirror fermions and a Higgs sector consisting of doublets and singlets are included. Parity can be broken at a much higher scale than the scale of new gauge boson. Charged fermion masses are generated from a seesaw mechanism, and the electron mass fixes the scale of one of the new scalars at the Peccei-Quinn scale. The new mirror neutrinos lead to a rich phenomenology. Two possibilities for the neutrino mass spectrum were presented in this work. New mirror neutrino can be very light. In this case, a new neutral gauge bososn can have a unique signature: a very high invisible branching ratio. The other possibility is the double seesaw mechanism. The high mass neutrino matrix has a second seesaw mechanism and we can have new neutrino states naturally at the TeV scale. The model presented in this work extends the heavy Majorana neutrino mass range to be searched for at the LHC  to the TeV region. The near start of the experimental program of the LHC will allow testing these hypothese.

\vfill\eject

\textit{Acknowledgments:} 
This work was supported by the following Brazilian agencies: CNPq, FAPERJ, FINEP and RENAFAE.

\end{document}